\documentclass[12pt]{article} 

\newcommand{\blind}{0}

\usepackage[
top    = 1.2in,
bottom = 1.4in,
left   = 1.0in,
right  = 1.0in]{geometry}
\usepackage{setspace}
\usepackage{times}
\usepackage{graphicx,psfrag,epsf}
\usepackage{amsmath}
\usepackage{amssymb}
\usepackage{gensymb}
\usepackage{subcaption}
\usepackage[toc,page]{appendix}
\usepackage{natbib}
\usepackage{xcolor}

\usepackage{graphicx} 

\usepackage{url}
\usepackage{booktabs}
\usepackage{array}
\usepackage{multirow}

\usepackage{enumitem}

\usepackage{verbatim}
\usepackage[format=plain, labelfont=bf, singlelinecheck=off]{caption} 

\usepackage{algorithm}
\usepackage{algpseudocode}
\newlength{\continueindent}
\usepackage{etoolbox}
\makeatletter
\newcommand*{\ALG@customparshape}{\parshape 2 \leftmargin \linewidth \dimexpr\ALG@tlm+\continueindent\relax \dimexpr\linewidth+\leftmargin-\ALG@tlm-\continueindent\relax}
\apptocmd{\ALG@beginblock}{\ALG@customparshape}{}{\errmessage{failed to patch}}
\makeatother

\numberwithin{equation}{section}

\newcommand{\bfu}{{u}}
\newcommand{\bfs}{{\mathbf{s}}}

\newcommand{\bfx}{{\mathbf{x}}}
\newcommand{\bfh}{{\mathbf{h}}}
\newcommand{\bfb}{{\mathbf{b}}}

\newcommand{\bfw}{{\mathbf{w}}}
\newcommand{\bfphi}{{\boldsymbol \phi}}

\newcommand{\bftheta}{{\boldsymbol \theta}}

\newcommand{\bfmu}{{\boldsymbol \mu}}
\newcommand{\bfSigma}{{\boldsymbol \Sigma}}

\newcommand{\bfD}{{\mathbf{D}}}

\newcommand{\bfH}{{\mathbf{H}}}

\newcommand{\bfV}{{\mathbf{V}}}
\newcommand{\bfR}{{\mathbf{R}}}

\newcommand{\bfY}{{\mathbf{Y}}}
\newcommand{\bfZ}{{\mathbf{Z}}}

\newcommand{\calX}{{\mathcal{X}^*}}

\newcommand{\bx}{{\mathbf{x}}}
\newcommand{\bb}{{\mathbf{b}}}

\newcommand{\bw}{{\mathbf{w}}}

\newcommand{\bZ}{{\mathbf{Z}}}
\newcommand{\bH}{{\mathbf{H}}}
\newcommand{\bI}{{\mathbf{I}}}

\expandafter\def\expandafter\normalsize\expandafter{%
    \normalsize
    \setlength\abovedisplayskip{6pt}
    \setlength\belowdisplayskip{6pt}
    \setlength\abovedisplayshortskip{6pt}
    \setlength\belowdisplayshortskip{6pt}
}

\usepackage{mathtools}

\begin{document}
\date{}

\def\spacingset#1{\renewcommand{\baselinestretch}%
{#1}\small\normalsize} \spacingset{1}


\if0\blind
{
  \title{\bf An Additive Approximate Gaussian Process Model for Large Spatio-Temporal Data
  }

  \author{Pulong Ma\footnote{ \emph{Correspondence to}: Pulong Ma, The Statistical and Applied Mathematical Sciences Institute and Duke University, 79 T.W. Alexander Drive, P.O. Box 110207, Durham, NC 27709. Email: pma@samsi.info} \\
 Statistical and Applied Mathematical Sciences Institute  \\
  and Duke University \\
    Bledar A. Konomi and Emily L. Kang \\
    University of Cincinnati\\
    }

  \maketitle

} \fi

\if1\blind
{
  \bigskip
  \bigskip
  \bigskip
  \begin{center}
    {\LARGE\bf An Additive Approximate Gaussian Process Model for Large Spatio-Temporal Data}
\end{center}
  \medskip
} \fi

\bigskip
\begin{abstract}

Motivated by a large ground-level ozone dataset, we propose a new computationally efficient additive approximate Gaussian process. The proposed method incorporates a computational-complexity-reduction method and a separable covariance function, which can flexibly capture various spatio-temporal dependence structure. The first component is able to capture nonseparable spatio-temporal variability while the second component captures the separable variation. Based on a hierarchical formulation of the model, we are able to utilize the computational advantages of both components and perform efficient Bayesian inference. To demonstrate the inferential and computational benefits of the proposed method, we carry out extensive simulation studies assuming various scenarios of underlying spatio-temporal covariance structure. The proposed method is also applied to analyze large spatio-temporal measurements of ground-level ozone in the Eastern United States.

\end{abstract}

\noindent%
\vfill

\newpage
\doublespacing

\section{Introduction}
\label{sec: intro}

In the United States, the Clean Air Act requires the Environmental Protection Agency (EPA) to set National Ambient Air Quality Standards (NAAQS) for pollutants that are considered harmful to public health and environment. EPA has set NAAQS for six principal pollutants: Carbon monoxide, lead, nitrogen dioxide, ozone, particulate pollution, and sulfur dioxide, which are called  ``criteria'' air pollutants. These standards are periodically reviewed and are subject to revision. The NAAQS for ground-level ozone (O$_3$) is set up based on the fourth-highest daily maximum 8-hour average ozone level over an ozone season. Because actual observations of ozone are sparse in space and irregular in time, it is crucial to develop statistical models that are able to make spatio-temporal predictions of ground-level ozone with quantified uncertainty. However, spatio-temporal modeling for these datasets can be challenging due to the presence of  nonseparability and high-dimensionality \citep{Fuentes2003,Gilleland2005}. 

To overcome computational challenges, many {computational-complexity-reduction} (CCR) methods have been proposed to analyze large or massive spatial/spatio-temporal data, including predictive process \citep[PP;][]{Banerjee2008}, modified predictive process \citep[MPP;][]{Finley2009}, dynamic nearest neighbor Gaussian process \citep{Datta2016}, and full-scale approximation \citep[FSA;][]{Sang2012,Zhang2015}. These methods have been developed in a continuously-indexed spatio-temporal domain, and can allow for predictions at arbitrary spatial locations and time points. They take advantage of a low-rank model, a low-order conditioning set, or a sparse covariance/precision matrix to alleviate computational difficulties of Gaussian process modeling. Indeed, they can be unified into a general framework via Vecchia approximations \citep{Vecchia1988} as discussed in \cite{Katzfuss2018}.  Specifically, the PP explains mainly the large-scale variation by a low-rank component but could result in biased parameter estimation as well as over-smoothed predictions of the spatial/spatio-temporal field. The MPP was introduced to deal with  positive bias in the non-spatial error term of the models in the PP.  The spatio-temporal FSA approach combines a reduced-rank covariance approximation with a tapering \citep{Furrer2006} or block covariance approximation. It can effectively capture both the large-scale and small-scale spatial/spatio-temporal variation. However, it can be computationally challenging for FSA to simultaneously find optimal number of knots and tapering range in real applications. The dynamic nearest neighbor Gaussian process was proposed to generalize the nearest neighbor Gaussian process \citep[NNGP;][]{Datta2015} in a spatio-temporal setting based on a dynamic selection of neighbor sets. This dynamic selection procedure was illustrated with a naturally monotonic space-time covariance function by fixing interaction parameter in the Gneiting's nonseparable covariance function, which may not always be a realistic assumption for modeling environmental processes \citep{Cressie2011}.

For ground-level ozone, the network of monitoring sites is typically fixed across time. This allows convenient spatio-temporal modeling based on a separable space-time covariance function. In particular, with a separable space-time covariance function, the resulting covariance matrix can be expressed as a Kronecker product of a purely spatial covariance matrix and a purely temporal covariance matrix. Thus, it can alleviate computational burdens for likelihood evaluation and model fitting \citep{Genton2007, Rougier2008, Bilionis2013}. Although a space-time separable covariance function could be favored in terms of computational efficiency, it is not appropriate to assume such separable dependence structure solely when modeling the ground-level ozone, since the ground-level ozone data exhibit space-time interactions, see illustrations in Section~\ref{sec: application} and the work in \cite{Gilleland2005} and \cite{Zhang2015}.

Motivated by the nonseparable dependence structure and the network structure of the ground-level ozone data, we propose an additive approximate Gaussian process (AAGP). The proposed AAGP consists of two independent and computationally efficient spatio-temporal Gaussian processes. In particular, the first component is modeled as an approximation of a nonseparable space-time process using a CCR method aforementioned. For simplicity, we choose to work with the MPP instead of the space-time FSA or the dynamic NNGP. The second component is assumed to be a separable space-time  process that is computationally favored by the network structure of monitoring sites for ground-level ozone. Since the variance parameters in these two Gaussian processes are allowed to be different, the overall covariance function in the AAGP can be viewed as a weighted average of the covariances from these two components, where the weights are determined by the proportion of their corresponding variances. The proposed model not only captures the nonseparable dependence structure that is approximated by a CCR method, but also is able to allow model selection when the data exhibit a separable dependence structure.

The reminder of this paper is organized as follows. Section~\ref{sec: AAGP} presents the basic definition of the additive approximate Gaussian process and its covariance function specification. In Section~\ref{sec: AAGP for large data}, we develop a fast Metropolis-within-Gibbs sampler based on a fully hierarchical formulation of AAGP, and also include remarks on alternative covariance specifications in AAGP. Section~\ref{sec: numerical illustration} demonstrates the predictive performance of AAGP with several simulation examples. In Section~\ref{sec: application}, we analyze ground-level ozone in the Eastern United States with the proposed method. Section~\ref{sec: conclusion} concludes with discussion on possible extensions of AAGP in future work.

\section{Additive approximate Gaussian process} \label{sec: AAGP}

Let $\{Z(\bfx): \bfx \in \mathcal{X} \equiv \mathcal{S} \times \mathcal{T}\}$ be a continuously-indexed spatio-temporal process, where $\bfx\equiv (\bfs, u)$ with $\bfs \in \mathcal{S}$ and $u\in \mathcal{T}$. Here, $\mathcal{S}\subset \mathbb{R}^d$ is a $d$-dimensional spatial domain with positive integer $d$, and $\mathcal{T}\subset \mathbb{R}$ is a temporal domain. Suppose that the spatio-temporal process $Z(\cdot)$ is observed at a total of $n$ locations, $\bfx_1, \ldots, \bfx_n \in \mathcal{X}$. We assume the following model for $Z(\cdot)$:
\begin{eqnarray} \label{eqn: data model}
Z(\bfx) = Y(\bfx) + \epsilon(\bfx),\, \bfx \in \mathcal{X},
\end{eqnarray}
where $Y(\cdot)$ is a latent Gaussian process of interest. The second term in the right-hand side of (\ref{eqn: data model}) is assumed to be  a Gaussian white-noise process with variance $\tau^2$, which is usually called the nugget effect. This term is commonly used to represent measurement errors for environmental data \citep{Cressie1993}. 

The process $Y(\cdot)$ is usually assumed to have additive components:
\begin{eqnarray} \label{eqn: process model}
Y(\bfx)=\bfh(\bfx)^T\bfb + w(\bfx),\, \bfx \in \mathcal{X},
\end{eqnarray}
where
$\bfh(\cdot)=[h_1(\cdot), h_1(\cdot), \ldots, h_{p}(\cdot)]^T$ is a vector of $p$ covariates; $\bfb$ is the corresponding vector of $p$ regression coefficients; $w(\cdot)$ is a Gaussian process with mean zero and covariance function $C(\cdot, \cdot)$. To allow efficient computation and to increase flexibility in the covariance structure, we assume that the process $w(\cdot)$ is approximated by a summation of two computationally efficient components $w_1(\cdot)$ and $w_2(\cdot)$. In particular, we assume that $w_1(\cdot)$ and $w_2(\cdot)$ are independent Gaussian processes with covariance functions $C_1(\cdot, \cdot)$ and $C_2(\cdot, \cdot)$, which come from two different covariance families. We call the resulting process $Y(\cdot)$ the additive approximate Gaussian process (AAGP). Its covariance function can be written as $\text{cov}(Y(\bfx), Y(\bfx')) = C_1(\bfx, \bfx') + C_2(\bfx, \bfx')$ with the two components $C_1(\cdot,\cdot)$ and $C_2(\cdot, \cdot)$ described below. 

In this paper we concentrate on two specific forms of covariance functions. We choose a nonseparable covariance function $C_1(\cdot,\cdot)$ to model potential spatio-temporal interaction. However, this type of covariance functions is computationally challenging when analyzing large spatio-temporal datasets. We thus use a CCR method to approximate $C_1(\cdot, \cdot)$. For convenience, the modified predictive process model is used to approximate $C_1(\cdot, \cdot)$ in this paper. The covariance function $C_2(\cdot, \cdot)$ is assumed to be a separable covariance function. This is mainly motivated by the network structure of the data, and the fact that a separable covariance function can capture all scales of variation. Notice that the choice of a nonseparable covariance function and a separable covariance function can avoid the non-identifiability issue since these two covariance functions characterize different dependence structures. 

Our method differs from previous methods that use an additive structure resulting from two different covariance components. For instance, \cite{Rougier2008} use a low-rank component plus a separable covariance function while the low-rank component is constructed with pre-specified regressors of input/output variables in a separable form. Our method explicitly includes the nonseparable dependence structure which is not necessarily low-rank. In addition, our model includes a nugget term recommended in modeling environmental data \citep{Cressie1993}, to ensure computational stability and better predictive performance. \cite{Ma2017} propose a model with a low-rank component and a Gaussian graphical model that induces a sparse precision matrix, but their method applies to spatial data instead of spatio-temporal data. \cite{Ba2012} use a sum of two independent GPs with separable squared exponential covariance functions to approximate computer model outputs, but they have to impose empirical constraints on parameters in these two covariance functions to avoid non-identifiability. Our method avoids such an issue by using different types of covariance structures and is designed to handle large datasets in a Bayesian framework.

The covariance structure in AAGP is fundamentally different from methods such as FSA and multi-resolution approximation \citep[MRA;][]{Katzfuss2016}, because those methods are designed to use multiple components altogether to approximate a target covariance function. In fact, both FSA and MRA are alternative CCR methods to model the component $w_1(\cdot)$ in AAGP.

\subsection{A computational-complexity-reduction covariance function}
To handle large data size $n$, we adopt an approximation method to reduce computational complexity for the component $w_1(\cdot)$ with a nonseparable covariance function $C_1(\cdot,\cdot)$. Predictive process\ methods \citep[e.g.,][]{Banerjee2008, Finley2009} have been proposed and applied successfully with large data. These methods use low-rank representations to allow reduced dimension and only require linear computational cost to invert or factorize large covariance matrices via the Sherman-Morrison-Woodbury formula. A brief review of this method is as follows. Suppose that a nonseparable correlation function $R_0(\cdot, \cdot; \bftheta_1)$ is known up to a few parameters $\bftheta_1$. To reduce dimensionality of the problem, only a pre-specified set of $m$ ($m\ll n$) knots $\mathcal{X}^*\equiv \{ \mathbf{x}_1^*, \ldots, \mathbf{x}_m^*\}\subset \mathcal{X}$ is chosen to project the original process $w_1(\cdot)$ to the space spanned by the collection of variables $\{w_1(\bfx_1), \ldots, w_1(\bfx_m)\}$. Specifically, the process $w_1(\cdot)$ is modeled as a GP with mean zero and correlation function given by:
\begin{eqnarray}
{R}_1(\bfx, \bfx') = \bfR(\bfx, \calX)\bfR_*^{-1}\bfR(\bfx', \calX)^T  + I(\bfx=\bfx')[1- \bfR(\bfx, \calX)\bfR_*^{-1}\bfR(\bfx', \calX)^T],
\end{eqnarray}
where $\bfR(\bfx, \calX)\equiv [R_0(\bfx, \bfx_i^*)]_{i=1, \ldots, m}$ is an $m$-dimensional row vector; $\bfR_*$ is the $m$-by-$m$ matrix with its $(i,j)$-th element $R_0(\bfx_i^\ast, \bfx_j^\ast)$, for $i,j=1,\ldots, m$, and  ${I}(\cdot)$ denotes the indicator function of its argument.  It is straightforward to show that $R_1(\bfx, \bfx') = 1$ if $\bfx=\bfx'$. Based on this construction, the correlation matrix of  $\bfw_1\equiv(w_1(\bfx_1), \ldots, w_1(\bfx_n))^T$ is $\mathbf{R}_1\equiv \mathbf{R}_{nm} \mathbf{R}^{-1}_* \mathbf{R}_{nm}^T + \mathbf{V}$, where $\bfR_{nm}\equiv [R_0(\bfx_i, \bfx_j^*)]_{i=1, \ldots, n; j=1, \ldots, m}$, and $\bfV$ is an $n$-by-$n$ diagonal matrix with its $i$th diagonal element given by $V_i \equiv 1- \bfR(\bfx_i,\calX)\bfR_*^{-1} \bfR(\bfx_i,\calX)^T$. Note that the vector $\bfR(\bfx,\calX)$ and matrices $\bfR_*$ and $\mathbf{V}$ all depend on the unknown parameters $\bftheta_1$. The resulting covariance function $C_1(\cdot, \cdot; \bftheta_1)$ is $\sigma^2_1R_1(\cdot, \cdot; \bftheta_1)$, where $\sigma^2_1$ is the variance parameter. Readers are referred to \cite{Finley2009} for more detailed model formulation and development. Although we use the MPP to approximate the nonseparable covariance function $C_1(\cdot,\cdot)$, we discuss in Section~\ref{sec: AAGP for large data} how the inference framework can be applied when alternative CCR methods are used to approximate the component $w_1(\cdot)$ with a nonseparable covariance function $C_1(\cdot,\cdot)$. 

\subsection{A separable covariance function} 
We assume a separable covariance function for the second component $w_2(\cdot)$. The benefits are two-fold: First, it enables the resulting model to flexibly model both nonseparable or separable processes (and the combination of both); secondly, the separable component can contribute to modeling the remaining variability due to the approximation in the CCR method.

For $\mathbf{s}, \bfs' \in \mathcal{S}, u, \bfu' \in \mathcal{T}$,  the process $w_2(\cdot)$ is assumed to have the variance parameter $\sigma^2$ with a separable correlation function: 
\begin{eqnarray} \label{eqn: separable model}
R_2(\bfx, \bfx'; \bftheta_2)= \rho_1(\mathbf{s}, \mathbf{s}'; \bfphi_1)\rho_2(u, u'; \bfphi_2),
\end{eqnarray}
where $\rho_1(\cdot, \cdot)$ and $\rho_2(\cdot, \cdot)$ are correlation functions with range parameters $\bfphi_1$ and $\bfphi_2$ over space $\mathcal{S}$ and $\mathcal{T}$, respectively. Let $\bftheta_2\equiv\{ \bfphi_1, \bfphi_2\}$ be a vector containing these range parameters. The data process $Z(\cdot)$ is assumed to be observed at all the $n=n_1n_2$ locations arranged as $\bfx_1=(\bfs_1, \bfu_1), \ldots, \bfx_{n_2}=(\bfs_1, \bfu_{n_2})$, $\bfx_{n_2+1}=(\bfs_2, \bfu_1), \ldots, \bfx_{2n_2}=(\bfs_2, \bfu_{n_2}), \ldots, \bfx_n=(\bfs_{n_1}$, $\bfu_{n_2})$, where $n_1$ denotes the number of spatial locations in $\mathcal{S}$, and $n_2$ denotes the number of time points in $\mathcal{T}$. The resulting correlation matrix of $\mathbf{w}_2\equiv(w_2(\bfx_1), \ldots, w_2(\bfx_n))^T$ is $\mathbf{R}_2 \equiv \mathbf{R}_s \otimes \mathbf{R}_u$, where $\mathbf{R}_s\equiv [\rho_1(\mathbf{s}_i, \mathbf{s}_j)]_{i,j=1, \ldots, n_1}$ is an $n_1$-by-$n_1$ matrix, and $\mathbf{R}_u \equiv [\rho_2(u_i, u_j)]_{i,j=1, \ldots, n_2}$ is an $n_2$-by-$n_2$ matrix. Notice that the locations $\{\bfs_1, \ldots, \bfs_{n_1}\}$ and $\{\bfu_1, \ldots, \bfu_{n_2}\}$ are not necessarily regularly spread out in $\mathcal{S}$ and $\mathcal{T}$. As shown in \cite{Genton2007} and \cite{Rougier2008}, imposing separability on the covariance function enables us to use attractive properties of Kronecker product of matrices, which brings substantial computational gains. The tentative assumption that  $Z(\cdot)$ is observed at all the $n=n_1n_2$ locations will be relaxed in Section~\ref{sec: missing data imputation}. We will illustrate there how a step of missing data imputation is added and embedded in Bayesian inference. In addition, we focus on the problem that $n$ is large (in order of $10^4\sim 10^6$) but $n_1$ and $n_2$ on their own are not very large (about or less than $10^3$). In Section~\ref{sec: conclusion}, several modeling strategies are recommended on how to extend the proposed method when either $n_1$ or $n_2$ is large.

\subsection{Likelihood evaluation}
Let $\mathbf{Z}\equiv (Z(\bfx_1), \ldots, Z(\bfx_n))^T$ be the vector of $n$ observations. Given the model specification in Equations~\eqref{eqn: data model} to \eqref{eqn: separable model}, the log-likelihood function of the data vector $\bZ$ can be written as
\begin{eqnarray} \label{eqn: marginal data model}
\ell(\bb, \tau^2, \sigma^2_1, \sigma_2^2, \bftheta_1, \bftheta_2;\bZ)= -n\log(2\pi)/2  
 - (\bZ-\bH\bb)^T\bfSigma^{-1}(\bZ-\bH\bb)/2, 
\end{eqnarray}
where $\mathbf{H}\equiv [\mathbf{h}(\bfx_1), \ldots, \mathbf{h}(\bfx_n)]^T$ is a matrix of covariates or regressors. $\bfSigma$ is the covariance matrix  of $\bZ$ with the following form
\begin{eqnarray} \label{eqn: cov matrix of data} 
\bfSigma &\equiv& \text{cov}(\mathbf{Z}) =\sigma^2_1\bfR_1 + \sigma^2_2\bfR_s\otimes \bfR_u + \tau^2 \mathbf{I}, \\ \nonumber
&=& \sigma^2_1 \bfR_{nm} \bfR_*^{-1} \bfR_{nm}^T +  \sigma^2_2\bfR_s\otimes \bfR_u + \sigma^2_1\bfV + \tau^2 \mathbf{I}.
\end{eqnarray}
Evaluation of this log-likelihood function involves the inversion and determinant of the $n$-by-$n$ covariance matrix $\bfSigma$. When $n$ is large, techniques such as the Sherman-Morrison-Woodbury formula and the Cholesky decomposition of sparse matrices are widely used to reduce computational complexity \citep[e.g.,][]{Banerjee2008, Cressie2008, Sang2012, Datta2015}. However, these techniques cannot be directly applied to AAGP to reduce computational complexity, as we will explain below, and we show the additive structure of AAGP requires careful handling, and we propose a fully conditional approach for its Bayesian inference. 

To simplify notations, we use $\bfD$ to denote the matrix $\sigma^2_2\bfR_s\otimes \bfR_u + \sigma^2_1\bfV + \tau^2 \mathbf{I}$. Then the Sherman-Morrison-Woodbury formula can be used to derive the formula for $\bfSigma^{-1}$:
\begin{eqnarray*}
\bfSigma^{-1}=\bfD^{-1} - \bfD^{-1} \bfR_{nm}(\sigma^{-2}_1\bfR_{*} 
+ \bfR_{nm}^T \bfD^{-1} \bfR_{nm})^{-1} \bfR_{nm}^T \bfD^{-1},
\end{eqnarray*}
where the inversion of $\bfD$ is required in order to solve linear systems involving $\bfSigma$. It is worth noting that calculating this inversion $\bfSigma^{-1}$ is not computationally feasible for large $n$. In particular, it requires inversions of two $m$-by-$m$ matrices, $\bfR_*$ and $\sigma^{-2}_1\bfR_{*} + \bfR_{nm} \bfD^{-1} \bfR_{nm}$, and inversion of the $n$-by-$n$ matrix $\bfD$. As $m$ is much smaller than $n$, inverting the $m$-by-$m$ matrices can be done easily with $O(m^3)$ flops, since $m$ is much smaller than $n$. However, inverting the $n$-by-$n$ matrix $\bfD$ requires full matrix inversion due to the presence of heterogeneous diagonal elements in $\sigma^2_1\bfV$ together with  $\sigma^2_2\bfR_s\otimes \bfR_u$.

Gaussian process regression is usually implemented via likelihood-based inference or fully Bayesian inference, which typically fits the marginalized model after integrating out random effects \citep[e.g.,][]{Ba2012, Banerjee2014}. Such inference procedures cannot be used when we fit AAGP due to the $O(n^3)$ computational cost to solve linear systems involving $\bfD$. To tackle this computational challenge, we propose a fully conditional Markov chain Monte Carlo (MCMC) algorithm in the next section. 

\section{Bayesian inference: a fully conditional approach} \label{sec: AAGP for large data}

To carry out Bayesian inference for AAGP, we first assign prior distributions to the unknown parameters $\{\bb, \tau^2, \sigma^2_1, \sigma_2^2, \bftheta_1, \bftheta_2\}$. Following customary prior specifications, we assign a vague multivariate normal prior for the coefficient vector $\mathbf{b} \sim \mathcal{N}_p(\bfmu_b, \mathbf{V}_b)$, independent inverse gamma priors for variance parameters: $\tau^2 \sim \mathcal{IG}(a_{\tau}, b_{\tau})$,  $\sigma^2_1 \sim \mathcal{IG}(a_1, b_1)$, $\sigma^2_2 \sim \mathcal{IG}(a_2, b_2)$, and independent uniform priors for other parameters in $\bftheta_1$ and $\bftheta_2$. 

Conventional fully Bayesian inference procedures for GP modeling typically focus on the marginal distribution of data after integrating out random effects. In the AAGP model, we can write out the (joint) posterior distribution $p(\mathbf{b}, \tau^2, \sigma^2_1, \sigma^2_2,$ $\bftheta_1, \bftheta_2 \mid \mathbf{Z})$, which is  proportional to the joint distribution:
\begin{eqnarray} \label{eqn: marginal joint posterior}
p(\bb, \tau^2, \sigma^2_1, \sigma_2^2, \bftheta_1, \bftheta_2)  p(\bZ|\bb, \tau^2, \sigma^2_1, \sigma_2^2, \bftheta_1, \bftheta_2)\hspace{1cm} & &  \\  \nonumber
=\mathcal{N}_p(\bfmu_b, \mathbf{V}_b)\, \mathcal{IG}(a_{\tau}, b_{\tau})
 \mathcal{IG}(a_1, b_1)  \mathcal{IG}(a_2, b_2) p(\bftheta_1, \bftheta_2) 
  \times \mathcal{N}_n(\mathbf{Hb}, \bfSigma). &&
\end{eqnarray}

Sampling from this posterior distribution~\eqref{eqn: marginal joint posterior} is computationally infeasible with large $n$,  since each MCMC iteration requires inversion of the $n$-by-$n$ covariance matrix $\bfSigma$, which requires $O(n^3)$ flops and $O(n^2)$ memory. Rather than utilizing the marginal distribution of $\bZ$, we write the model in a hierarchical form with the latent processes $w_1(\cdot)$ and $w_2(\cdot)$. This allows the development of a computationally efficient MCMC sampling procedure for fully Bayesian inference. 

The data model in Equation~\eqref{eqn: data model} and the process model in Equation~\eqref{eqn: process model} give a hierarchical formulation of the AAGP model:
\begin{eqnarray} \label{eqn: data model in stage}
\mathbf{Z} \mid \mathbf{b}, \mathbf{w}_1, \mathbf{w}_2, \tau^2 \, &\sim& \, \mathcal{N}_n (\mathbf{Hb} + \mathbf{w}_1 + \mathbf{w}_2, \tau^2\bI_n), \\
\label{eqn: process model 1 in stage}
\mathbf{w}_1\mid \sigma^2_1, \bftheta_1 &\sim& \mathcal{N}_n(\bfR_{nm} \bfR_*^{-1} \bfw^*, \sigma^2_1\bfV), \\
\label{eqn: process model 2 in stage}
\mathbf{w}_2\mid\sigma^2_2, \bftheta_2 &\sim& \mathcal{N}_n(\mathbf{0}, \sigma^2_2\bfR_s\otimes \bfR_u), 
\end{eqnarray}
where $\bfw^* \equiv (w_1(\bfx_1^*), \ldots, w_1(\bfx_m^*))^T$ is an $m$-dimensional random vector following the multivariate normal distribution with mean zero and covariance matrix $ \sigma^2_1\bfR_*$. The joint posterior distribution of unknown parameters $\{\mathbf{b}, \tau^2, \sigma^2_1, \sigma^2_2, \bftheta_1, \bftheta_2\}$ and latent random effects $\{\mathbf{w}^*$, $\mathbf{w}_1$, $\mathbf{w}_2\}$ can be obtained as follows: 
\begin{eqnarray} \label{eqn: joint posterior}
p(\mathbf{b},\tau^2, \sigma^2_1, \sigma^2_2, \bftheta_1, \bftheta_2 ,\mathbf{w}^*, \mathbf{w}_1, \mathbf{w}_2 \mid \mathbf{Z}) \hspace{1cm}  && 
\\ \nonumber
\propto \mathcal{N}_p(\bfmu_b, \mathbf{V}_b) \mathcal{IG}(a_{\tau}, b_{\tau})
 \mathcal{IG}(a_1, b_1)  \mathcal{IG}(a_2, b_2) p(\bftheta_1, \bftheta_2)  &&
 \\  \nonumber
 \times  \mathcal{N}_m(\bfw^*\mid \mathbf{0},   \sigma^2_1 \bfR_*) \times   \mathcal{N}_n(\bfw_1\mid\bfR_{nm} \bfR_*^{-1} \bfw^*, \sigma^2_1\bfV)  && \\ \nonumber
 \times   \mathcal{N}_n(\bfw_2\mid \mathbf{0}, \sigma^2_2 \bfR_s \otimes \bfR_u) \times \mathcal{N}_n(\bfZ\mid \bfH \bfb+ \bfw_1+\bfw_2, \tau^2\mathbf{I}_n). & &
\end{eqnarray}

\subsection{Parameter estimation \& computational cost} \label{sec: MHGS}
Since the posterior distribution~\eqref{eqn: joint posterior} is intractable, we use a Metropolis-within-Gibbs sampler \citep{Hastings1970, Gelfand1990} for parameter inference. In particular, the conjugate full conditional distributions for $\bb, \tau^2, \sigma^2_1, \sigma^2_2$, and multivariate normal full conditional distributions for random effects $\mathbf{w}^*$, $\mathbf{w}_1$, and $\mathbf{w}_2$ are available in closed-form. 
To sample $\bftheta_1$ and $\bftheta_2$ from their full conditional distributions, a Metropolis-Hastings step is incorporated for each parameter, since these full conditional distributions are not any standard distribution. The detailed sampling procedure is outlined in the Supplementary Materials.

The hierarchical formulation of the model leads to a computationally efficient Metropolis-within-Gibbs sampler. In terms of computational cost, sampling from the full conditional distributions of $\mathbf{b}$, $\tau^2$, $\sigma^2_1$, $\sigma^2_2$, $\mathbf{w}^*$, $\mathbf{w}_1$,  and $\mathbf{w}_2$ requires $O(m^3+m^2n+n_1^3+n_2^3+n(n_1+n_2))$ flops. Sampling from full conditional distributions for $\bftheta_1$ and $\bftheta_2$ requires $O(m^3+m^2n + n_1^3+n_2^3+n(n_1+n_2))$ flops. Therefore, the overall computational cost for each MCMC iteration is  $O(m^2n+n_1^3+n_2^3+n(n_1+n_2))$. Note that $m$, $n_1$, and $n_2$ are all smaller than $n$, which makes this inference procedure much more efficient than making inference based on the marginal distribution of the data.  Although we sample the $n$-dimensional vectors $\bfw_1$ and  $\bfw_2$ in the Gibbs sampler in each MCMC iteration, there is no need to store all samples of these two high-dimensional vectors, because they can always be recovered through $[\bfw_1|\bfZ]=\int [\bfw_1| \sigma^2_1, \bftheta_1] [ \sigma^2_1, \bftheta_1 | \bfZ] \,d\{\sigma^2_1, \bftheta_1\}$ and $[\bfw_2|\bfZ]=\int [\bfw_2| \sigma^2_2, \bftheta_2] [\sigma^2_2, \bftheta_2 | \bfZ] \,d\{\sigma^2_2, \bftheta_2\}$. Therefore, the overall memory cost for each MCMC iteration is roughly $O(mn+n_1^2+n_2^2)$.


\subsection{Prediction} \label{sec: prediction}

For any location $ \bfx_0=(\bfs_0, \bfu_0) \in \mathcal{X}$, our interest is to make prediction for $Y(\bfx_0)$. Define $\Omega=\{\bfb, \sigma^2_1, \sigma_2^2, \tau^2, \bftheta_1, \bftheta_2\}$. The (posterior) predictive distribution of $Y(\bfx_0)$ given $\bfZ$ is
\begin{eqnarray} \nonumber
p(Y(\bfx_0)\mid \bfZ) &=& \int p(Y(\bfx_0) \mid\bfw_1, \bfw_2, \Omega, \bfZ)
  p(\bfw_1, \bfw_2, \Omega \mid \bfZ) \text{ d}\{\bfw_1, \bfw_2, \Omega\} \\ \nonumber
&=& \int p(Y(\bfx_0) \mid\bfw_1, \bfw_2, \Omega) 
p(\bfw_1, \bfw_2, \Omega \mid \bfZ) \text{ d}\{\bfw_1, \bfw_2, \Omega\}.
\end{eqnarray}
Samples from the predictive distribution $p(Y(\bfx_0)\mid\bfZ)$ can be obtained using composition sampling technique. That is, we draw from $p(Y(\bfx_0) \mid \bfw_1, \bfw_2, \Omega)$, where $\bfw_1, \bfw_2, \Omega$ are draws from the posterior distribution $p(\bfw_1, \bfw_2, \Omega \mid \bfZ)$. The formula of the predictive distribution is given in the Supplementary Material. 

\subsection{Alternative specification}
The AAGP model relies on a CCR covariance function model and a separable covariance function model. The modified predictive process (MPP) is chosen to derive the CCR covariance function and to illustrate the computational benefit of the proposed fast Bayesian inference procedure. However, it should be noted that the proposed inference procedure still applies when we choose an alternative CCR method for the nonseparable component in AAGP. As recently noted in \cite{Katzfuss2018}, MPP is a special case of more general Vecchia approximations, which include other existing methods such as FSA, NNGP, and MRA. These methods can also be used to derive the CCR covariance function. The corresponding Bayesian inference still works for large datasets. In particular, the matrix $\bfV$ in Equation~\eqref{eqn: cov matrix of data} will be replaced by a sparse matrix when FSA is used. The proposed inference procedure described above can still be applied efficiently. For NNGP and MRA, the vector $\bw^\ast$ will be high-dimensional, because these two methods use a smaller number of conditioning set to construct a sparse precision matrix rather than resorting to a low-rank structure for the covariance matrix. Note that the resulting covariance matrix of $\bfw^\ast$ is a sparse matrix. The proposed inference procedure can thus be implemented efficiently.

\subsection{{Missing data imputation}} \label{sec: missing data imputation}

Recall that we represent $\mathcal{X}$ as a product space $\mathcal{S}\times \mathcal{T}$ and have tentatively assumed that the response $Z(\cdot)$ is observed at all $n=n_1n_2$ locations, where $n_1$ denotes the number of unique spatial locations in $\mathcal{S}$,  and $n_2$ denotes the number of unique time points in  $\mathcal{T}$. This assumption is rarely satisfied for environmental data. In this subsection, we relax this assumption and explain how missing data imputation can be carried out. To fix the notation, we use $\mathcal{D}_c\equiv\{(\bfs_i, \bfu_j): \bfs_i \in \mathcal{S}, \bfu_j \in \mathcal{T}, i = 1, \ldots, n_1; j=1,\ldots, n_2\}$ to denote the complete grid over $\mathcal{X}=\mathcal{S} \times \mathcal{T}$. We assume that the data process $Z(\cdot)$ is only observed at a subset of $n$ ($n< n_1n_2$) locations $\mathcal{D}_o\equiv \{\bx_1, \ldots, \bx_n\}\subset \mathcal{D}_c$. The resulting $n$-dimensional data vector is denoted by $\bZ_o$, and we let $\bfZ_{\text{m}}$ denote the $(n_1n_2-n)$-dimensional vector of $Z(\cdot)$ at the unobserved locations in $\mathcal{D}_{\text{m}}\equiv \mathcal{D}_c\setminus \mathcal{D}_o$. In the Metropolis-within-Gibbs sampler, we now use $\bw_1$, $\bw_2$, and $\bZ$ to represent the $(n_1n_2)$-dimensional vectors at all locations in $\mathcal{D}_o$, and treat $\bZ_m$ as unknown. The full conditional distributions and sampling procedure for parameters $\{\bb, \tau^2, \sigma^2_1, \sigma^2_2, \bftheta_1, \theta_2\}$ and random effects $\bw^\ast$, $\bw_1$ and $\bw_2$ are the same as described in Section~\ref{sec: MHGS}. The missing values $\bZ_\text{m}$ can also be easily updated in MCMC based on its full conditional distribution. Actually, it can be shown that  $[\bfZ_{m} \mid \cdot]=\mathcal{N}( \bfY_{\text{m}}, \tau^2 \mathbf{I})$, where $\bfY_{\text{m}}\equiv \bH_\text{m}\bb+\bw_{1,\text{m}} +\bw_{2,\text{m}}$ with $\bH_\text{m}$ being a matrix of covariates, $\bw_{1,\text{m}}$ and $\bw_{2,\text{m}}$ being subsets of the random effects $\bw_1$ and $\bw_2$ over the unobserved locations in $\mathcal{D}_\text{m}$, respectively.

\section{Numerical illustrations} \label{sec: numerical illustration}
This section presents three simulation examples to illustrate the model adequacy and predictive accuracy of the proposed method AAGP, which is compared with the modified predictive process and the nearest neighbor Gaussian process. In addition, the full Gaussian process, referred to as Full GP, is used as benchmark in all synthetic examples. All these methods are implemented in MATLAB R2015b on a 10-core HP Intel Xeon E5-2680 machine with 12 GB random-access memory. To compare each method, we use 2.5th, 50th, 97.5th percentiles of model parameters, mean-squared-prediction errors (MSPEs), and average length of 95\% credible intervals (ALCI) for predictive values, to assess model adequacy and predictive accuracy. The total computing time is also reported for each method.

The purpose of these simulation examples is to investigate whether the AAGP can offer any computational and inferential benefits over other methods such as MPP and NNGP when the underlying true fields show different types of spatio-temporal dependence structures. In Supplementary Material, we also include a simulation example to demonstrate the AAGP with a spatio-temporal field generated from a deterministic function. In all these numerical examples, we use a class of Gneiting's nonseparable correlation functions \citep{Gneiting2002}, since this type of correlation functions is easy to interpret and has been widely used to model space-time interaction. In particular, we use the following form of Gneiting's nonseparable correlation function
\begin{eqnarray} \label{eqn: Gneiting's cov}
\rho((\bfs, \bfu), (\bfs', \bfu')) =  \left(\frac{(\bfu-\bfu')^{2\alpha}}{a} + 1\right)^{-d/2}
\cdot \exp\left\{ - \frac{\|\bfs-\bfs'\|}{c(\frac{(\bfu-\bfu')^{2\alpha}}{a} + 1)^{\beta/2}} \right\}, 
\end{eqnarray}
where $d$ is the dimension of the spatial domain $\mathcal{S}$; $a$ is the temporal range parameter in $\mathcal{T}$; $c$ is the spatial range parameter in $\mathcal{S}$; $\alpha \in (0, 1]$ is the smoothness parameter in $\mathcal{T}$; $\beta \in [0, 1]$ is the interaction parameter between $\mathcal{S}$ and $\mathcal{T}$.

To demonstrate the inferential and computational benefit of the AAGP model, three different scenarios with different space-time covariance structures will be implemented for the simulated true field $Y(\cdot)$ in a spatio-temporal domain $\mathcal{X}\equiv [0, 20]^2\times[1, 20]$. Specifically, the following three scenarios for the underlying true field $Y(\cdot)$ are considered:  
\begin{itemize}[noitemsep, topsep=0pt]
\item[(1)] Gneiting's space-time nonseparable correlation function only, referred to as Scenario 1;  
\item[(2)] separable correlation function only, referred to as Scenario 2;  
\item[(3)] a combination of Gneiting's space-time nonseparable correlation function and separable correlation function, referred to as Scenario 3.
\end{itemize}
The covariates in trend term contain $h_1(\bfx)$ and $ h_2(\bfx)$, where $h_1(\bfx)$ is simulated from the standard normal distribution, and $h_2(\bfx)=\cos(\mathbf{1}^T\bfx)$ for $\bfx \in \mathcal{X}$. Then the true process $Y(\cdot)$ is simulated on 4500 randomly-selected locations in the spatio-temporal domain $\mathcal{X}=[0, 20]^2\times [0, 20]$ with 225 spatial locations and 20 time points. The data $\bfZ$ are obtained by adding measurement errors whose variance $\tau^2$ is 0.2 in all the three scenarios.  For all the simulated data, $90\%$ of them are randomly selected as training set for model fitting, and the remaining $10\%$ are held out to evaluate predictive performance. 

In each scenario, the following models have been implemented: Full GP, MPP, NNGP and AAGP, where the Full GP is served as a benchmark. MPP and NNGP are two instances of CCR methods. In all the three scenarios, we use a single target covariance function in MPP and NNGP, since one typically prespecifies a single target covariance function in these CCR methods, say Gneiting's nonseparable covariance function, based on exploratory analysis such as variogram estimation. Note that the MPP is a sub-model of the AAGP in current examples. So, adding a separable covariance function model in the AAGP can improve the performance. Previous work has shown that the NNGP gives better results than the MPP for spatial and spatio-temporal data \citep{Datta2015, Datta2016}. It is interesting to investigate whether the implementation of the AAGP can have good performance compared with the NNGP from a modeling perspective without considering the structure of the data. In our implementation, we use the NNGP with the Gneiting's nonseparable covariance structure. Although it is also possible to set the target covariance fucntion in the NNGP to be of the additive form, we found in our numerical studies that the MCMC algorithm for the NNGP does not converge under such an additive setting. Our conjecture is that limiting a small neighborhood structure in the conditional distribution may make it difficult to identify the two components in the additive covariance function.  

In our numerical studies, we implemented the AAGP with 250 knots. The MPP is implemented with 250 knots and 704 knots, respectively. With 704 knots, the MPP costs about the same amount of time as the AAGP does. The NNGP is implemented with 15 nearest neighbors using sequential update in the MCMC algorithm shown in \citep{Datta2016}. The reference set is chosen to be the set of all observation locations. For all these methods, independent customary prior distributions are assigned: (1) $\mathbf{b} \sim \mathcal{N}_2(\mathbf{0}, 1000\mathbf{I})$; (2) $\sigma^2_1\sim \mathcal{IG}(2, 0.01)$; (3) $a\sim U(0, 20)$; (4) $c\sim U(0, 20)$; (5) $\beta\sim U(0, 1)$; (6) $\tau^2\sim \mathcal{IG}(2, 0.01)$. The smoothness parameter $\alpha$ is fixed at $0.5$ in the Gneiting's correlation function. The prior distributions for parameters in the space-time separable covariance function are specified as: $\sigma^2_2\sim \mathcal{IG}(2, 0.01)$,  $\phi_s\sim U(0, 20)$, and $ \phi_t\sim U(0, 20)$. The MCMC algorithm is run with 25000 iterations for each method with a burn-in period of 15000 iterations indicating independence from standard convergence diagnostics. In addition, we also add very small fixed nuggets to spatial and temporal separable correlation matrices $\bfR_s$ and $\bfR_u$ to avoid numerical instabilities in the MCMC algorithm. 

\subsection{Simulation example with a nonseparable covariance function}
In Scenario 1, the latent true process $Y(\cdot)$ is assumed to have a Gneiting's space-time covariance function with their parameters specified in the second column of Table~\ref{table: scenario 1}. The posterior summaries based on all the methods, Full GP, MPP, NNGP, and AAGP, are reported in Table~\ref{table: scenario 1}. Note that when fitting the Full GP, we assume the correct covariance function. Therefore, as expected, Full GP gives the smallest MSPE, while AAGP gives the second smallest MSPE. Specifically, the AAGP gives better prediction results than the MPP and the NNGP, since the MSPE from the AAGP is more than 20\% smaller than that from the MPP and the NNGP. This indicates that the separable component in the AAGP can capture part of the unexplained variability from the MPP, since the separable model can capture all scales of variability, and the MPP only captures large-scale variability in general. The spread of predictive distribution is very similar for both MPP and AAGP, but the predictive distribution for the AAGP is slightly more accurate than that for the MPP, and slightly worse than that for the NNGP. 

In terms of parameter estimation, the regression coefficients $b_1, b_2$ in the AAGP are estimated very well in comparison to the results in Full GP. The posterior mean for the variance parameters are $\sigma^2_1=0.869$ and $\sigma^2_2=0.303$ for the first and second component, respectively. This shows a clear preference for the Gneiting's nonseparable covariance function, since the $\sigma^2_1$ is much large than $\sigma^2_2$. The proposed AAGP is able to automatically assign the variation missed by the low-rank component to the separable component and the nugget. We also notice that the nugget is under-estimated and the overall variance $\sigma^2_1+\sigma^2_2$ is overestimated. One possible explanation of the above value is that the sum of estimated $\sigma^2_2$ and $\tau^2$ both together play the role of ``nugget effect''.  It is worth noting that the space-time interaction parameter $\beta$ has very wide credible interval even in the Full GP model, which indicates that this parameter cannot be estimated accurately even under the true model. However, the percentage of the variance parameters in the two components of AAGP provides a new way to characterize the space-time interaction.

\subsection{Simulation example with a separable covariance function}
In Scenario 2, the latent true process $Y(\cdot)$ is assumed to have the squared exponential correlation functions in space and time with parameters specified in the second column of Table~\ref{table: scenario 2}. The goal of this example is to investigate whether AAGP can detect the separability and to compare its performance with other models. In terms of model adequacy and predictive performance, the AAGP gives better MSPE and ALCI than both MPP and NNGP. The prediction results in AAGP are very close to the results of Full GP. The MPP gives the worst performance among all the methods. Its performance doesn't improve even though more knots are added in the MPP. The NNGP gives much better result than  the MPP, but its performance is still far behind from either AAGP or Full GP.

Posterior summaries in Table~\ref{table: scenario 2} suggest that both MPP and NNGP fail to detect the separability of the true field with the Gneiting's space-time correlation function, since $\beta$ has 95\% credible interval spreading out almost its entire support $[0, 1]$. The failure of MPP and NNGP on detecting the separability may also be related to the fact the Gneiting's space-time correlation function is not as smooth as the process with the squared exponential correlation function. To improve results for MPP and NNGP, we have also tried to take $\alpha$ to be a random variable. However, this leads to computational instabilities in the MCMC algorithm.  In contrast, the AAGP can detect the separability, since the estimated variance parameter $\sigma^2_1$ is close to 0, and estimated variance parameter $\sigma^2_2$ is close to 1. The trend parameters $b_1, b_2$ can be estimated very well. We can see that the posterior mean of the variance parameter $\sigma^2_1$ is close to 0, and  the posterior mean of the variance parameter $\sigma^2_2$ is close to 1. These two variance parameters serve as weights for the two components in the AAGP, and they are correctly identified: $\sigma^2_1=0.03$ and $\sigma^2_2=0.996$. The results in the MPP also show that the MPP's performance can deteriorate seriously when the covariance function is misspecified. 


\subsection{Simulation example with an additive covariance structure}
In Scenario 3, we address the problem of parameter estimation and predictive performance in AAGP under a true covariance function model. The process $Y(\cdot)$ is simulated from an additive Gaussian process with a Gneiting's space-time covariance function and a separable squared exponential covariance function with parameters specified in the second column of Table~\ref{table: scenario 3}. Posterior summaries for each model are reported in Table~\ref{table: scenario 3}. In terms of predictive performance, the MSPE in the AAGP is 66\% smaller than that in the MPP. The predictive performance in the AAGP is relatively close to the predictive performance of Full GP. But the performance of the AAGP deviates from that in the Full GP. One reason for this is that the MPP with 250 knots is used as the CCR method in the AAGP, which is not enough to capture the variability in the data. One could also image that if the NNGP is used as a CCR method, the performance of AAGP would be much better, since the the NNGP gives much better prediction results than the MPP. 

In terms of parameter estimation, the posterior mean of $b_1, b_2, \beta, \sigma^2_2, \phi_s, \phi_t$ are well estimated in the AAGP. The variance parameter $\sigma^2_1$ and range parameter $a$ are slightly over-estimated. We also observe that the nugget is under-estimated in the AAGP. This is likely because fixed small constants $\tau^2_s, \tau^2_u$ are added to the diagonal of the separable correlation matrices $\bfR_s$ and $\bfR_u$. As a consequence, it makes the actual nugget term to be $(\tau^2 + \tau^2_s \tau^2_u)$ instead of $\tau^2$.  As the variance parameters $\sigma^2_1$ and $\sigma^2_2$ are correctly identified in the AAGP, the proposed model AAGP is able to determine the variations coming from the non-separable and separable part automatically. The range parameter $c$ in the AAGP is over-estimated, and this is likely due to the overestimation of the variance parameter $\sigma^2_1$, since their ratio plays an important role in predictions \citep[for details, see][]{Kaufman2013}.

To briefly summarize our findings from these simulation examples, we found that the AAGP can give better prediction results than the MPP even though more knots are included in the MPP. This indicates that by adding an additional separable covariance function model, the AAGP outperforms its CCR method alone although we did not implement AAGP with all other CCR methods such as NNGP. The NNGP is a very appealing approach based on an attractive model development. However, the NNGP is not be able to give better prediction results than the AAGP in our simulation examples. It cannot detect the space-time interaction, especially in Scenario 2. So adding a separable covariance function in the AAGP brings inferential benefits. Our current implementation of the AAGP can be extended to incorporate other CCR methods such as NNGP. This will even improve the performance of the AAGP, since NNGP is known to perform better than MPP.

 With different underlying true covariance structures, the AAGP is able to give more robust prediction results than MPP and NNGP under misspecified covariance function models. This is crucial for spatio-temporal modeling in real applications. The inference procedure in the AAGP provides a computationally efficient strategy to allow fast Bayesian inference when a CCR method and a separable covariance function model are combined. It is worth mentioning that knots are selected uniformly in the MPP, and more sophisticated way to select the knots in the MPP is beyond the scope of this paper, for details, see \cite{Guhaniyogi2011}. The neighbors in the NNGP are chosen based on 15 nearest reference locations with the reference set chosen to be the set of observation locations. These implementation can be tuned to improve the performance of both MPP and NNGP. But it does not affect our conclusion on AAGP, since the AAGP can be built on CCR methods including MPP and NNGP.

\section{Analysis of Eastern US ozone data} \label{sec: application}

Ground-level ozone ($\text{O}_3$) is one of six common air pollutants identified in the Clean Air Act. To protect human health and the environment, EPA publishes the National Ambient Air Quality Standards (NAAQS) for ozone, which specifies the maximum allowed measurement for ozone to be present in the outdoor air. The NAAQS for ozone is calculated based on the following steps: 1) the maximum 8-hour average is calculated for each day; 2) then the fourth-highest value is computed for these daily maximum 8-hour averages; 3) finally, the NAAQS for ozone is defined as the average of these fourth-highest values for any consecutive three-year period. The proposed method is illustrated with daily maximum 8-hour average data at a network of monitoring sites in the Eastern U.S. from April through October in the year from 1995 to 1999. This data has been widely used in environmental statistics \citep[see, for example, ][]{Fuentes2003,Gilleland2005, Zhang2015}, and can be obtained from the website at \url{https://www.image.ucar.edu/Data/Ozmax}. Following the pre-processing steps in \cite{Gilleland2005}, the daily maximum 8-hour ozone average with unit parts per billion (ppb) at station $\bfs$ and day $u$, denoted by $O(\bfs, u)$, is assumed to have the following structure
\begin{eqnarray*}
O(\bfs, u) = \mu(\bfs, u) + \widetilde{O}(\bfs, u),  
\end{eqnarray*}
where $\mu(\bfs, u)=a(\bfs) + \sum_{j=1}^3\{ b_j \cos(2\pi ju/184) + c_j\sin(2\pi ju/184)\}$, which models the seasonal effect. The coefficients in the seasonal effect $\mu(\bfs, u)$ are estimated through ordinary least square method. The spatial-varying standard deviation $k(\cdot)$ is estimated based on residuals after removing the seasonal effect. 
The residual $r(\bfs, u)\equiv O(\bfs, u) - \hat{\mu}(\bfs, u)$ scaled by its estimated standard deviation $\hat{k}(\cdot)$ at each station is referred to as \emph{standardized ozone} at station $\bfs$ and time $u$ hereafter. 

\begin{figure*}[htbp]
\begin{center}
\makebox[\textwidth][c]{ \includegraphics[width=1.0\textwidth, height=0.5\textheight]{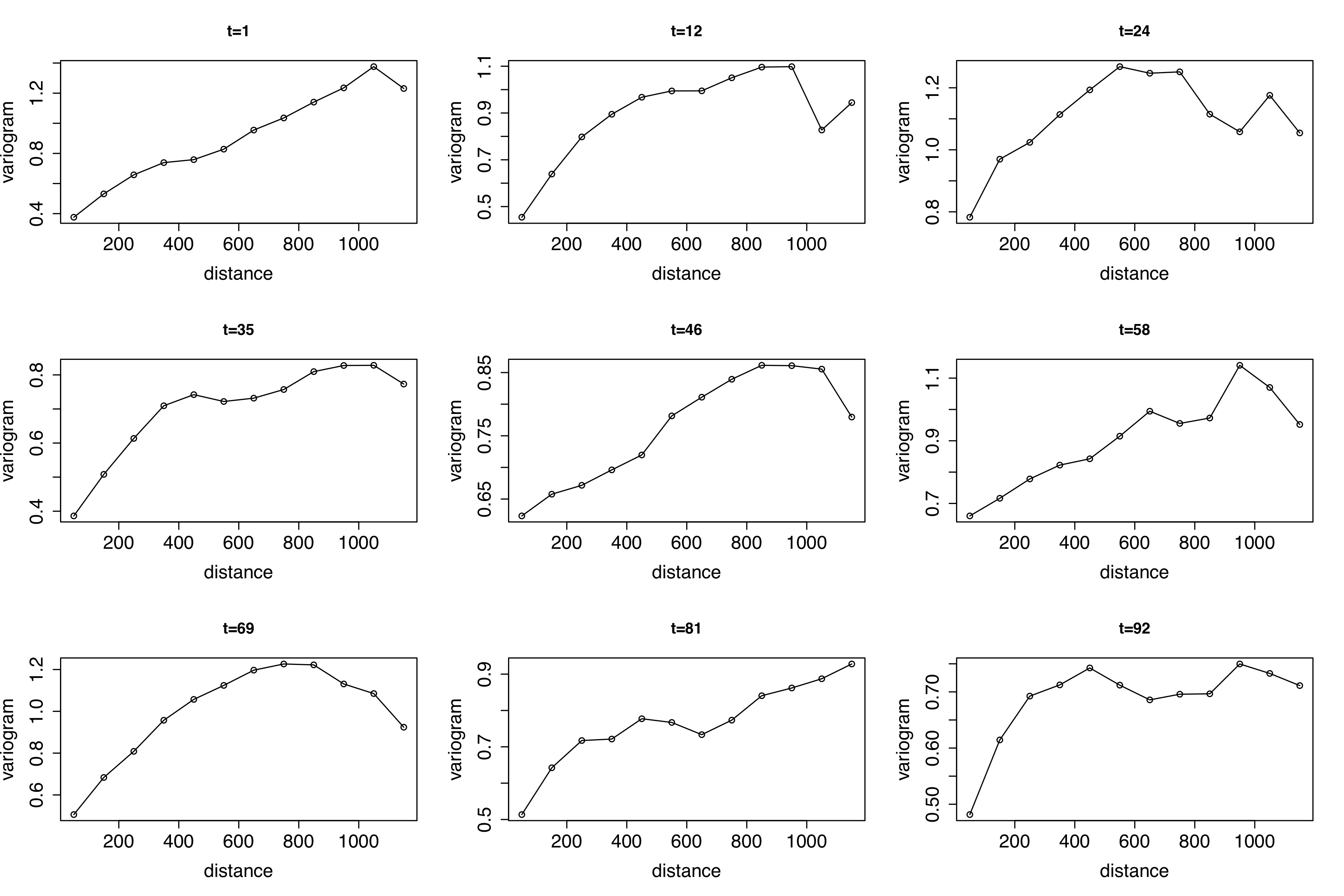}}
\caption{Empirical variograms of standardized ozone at nine different days.}
\label{fig: variogram}
\end{center}
\end{figure*}

The empirical variograms in Figure~\ref{fig: variogram} show that the spatial dependence structure of standardized ozone varies at 9 different time points. This suggests a nonseparable spatio-temporal covariance function model, since the spatial dependence structures are different across different time points. This illustration is consistent with the findings in \cite{Gilleland2005} and \cite{Zhang2015}. We also found that the empirical variograms at all time points give the sill around 1 and range less than 1200 kilometers. This information is used to setup the prior distributions in AAGP.

We perform the statistical analysis on the datasets collected at 513 monitoring sites during 92 days from June to August in 1997, where $1.37\%=645/(513\times 92)\times 100\%$ of data are missing, and only 46551 data points are observed. To analyze these data, a cross-validation procedure is first carried out on 46551 data points, where $90\%$ randomly selected data points are used for parameter estimation, and the remaining 10\% data points are held out to assess predictive performance. In the cross-validation, three methods are compared: MPP and NNGP with Gneiting's space-time covariance function, and AAGP with Gneiting's space-time covariance function and exponential covariance functions in the separable covariance function. Based on exploratory analysis, the prior distributions are specified as $\sigma^2_1\sim \mathcal{IG}(2, 0.01), \sigma^2_2\sim \mathcal{IG}(2, 0.01), \tau^2\sim \mathcal{IG}(2, 0.01), a\sim U(0, 60), c\sim U(0, 2000), \beta\sim U(0, 1)$ in Gneiting's space-time correlation function and $\phi_s\sim U(0, 2000), \phi_u\sim U(0, 60)$ in separable covariance functions in space and time. In MPP and AAGP, $490$ knots are selected in the spatio-temporal domain via Latin hypercube design. Then we further increase the number of knots up to 1200 in MPP to investigate whether the AAGP with just 490 knots still outperforms the MPP. The distance in space is calculated based on chordal distance, and the distance in time is calculated based on Euclidean distance. The NNGP model is implemented with 15 nearest neighbors with reference set being the set of observation locations. 

The posterior summaries in Table~\ref{table: cross validation for Jun-Aug} show that the AAGP gives better prediction results than the MPP even though more knots are added. The estimated variance for the Gneiting's nonseparable covariance function in the MPP is much larger than the variance estimated in the AAGP. The overall variance is estimated consistently based on MPP and AAGP. The NNGP also gives slightly larger MSPE and ALCI than AAGP. This indicates that the predictive distribution of AAGP is slightly more accurate than that in NNGP. As the standardized ozone data at each time point has variance around 1. Both NNGP and AAGP gives very good results. The interpolation of these ozone data hence can be reliable. 

The computing time for AAGP is roughly twice the computing time for MPP with the same number of knots. The AAGP is much faster than the NNGP model in a same software platform. It is worth mentioning that constructing a covariance matrix is slow and unavoidable in all these models. The construction of the correlation matrices $\bfR_1$ and $\bfR_*$ takes about 30\% of the total time in one MCMC iteration in the MPP, since these matrices need to be evaluated five times for one MCMC iteration. This unavoidable computing time can potentially make MPP as well as AAGP slow for very large datasets. For the NNGP model, as noted in \cite{Finley2018} that sequential updating MCMC algorithm can be very slow, but implementation of more efficient MCMC algorithms is beyond the scope of this paper. As we implemented all these methods in MATLAB, more speedups can be obtained if they were programmed in low-level languages such as C++. We would also like to point out that the CCR method in AAGP is derived from MPP, but other methods such as NNGP or MRA can also be used in AAGP to achieve further inferential benefit. 

\begin{figure*}[htbp]
\begin{center}
\makebox[\textwidth][c]{ \includegraphics[width=1.0\textwidth, height=0.6\textheight]{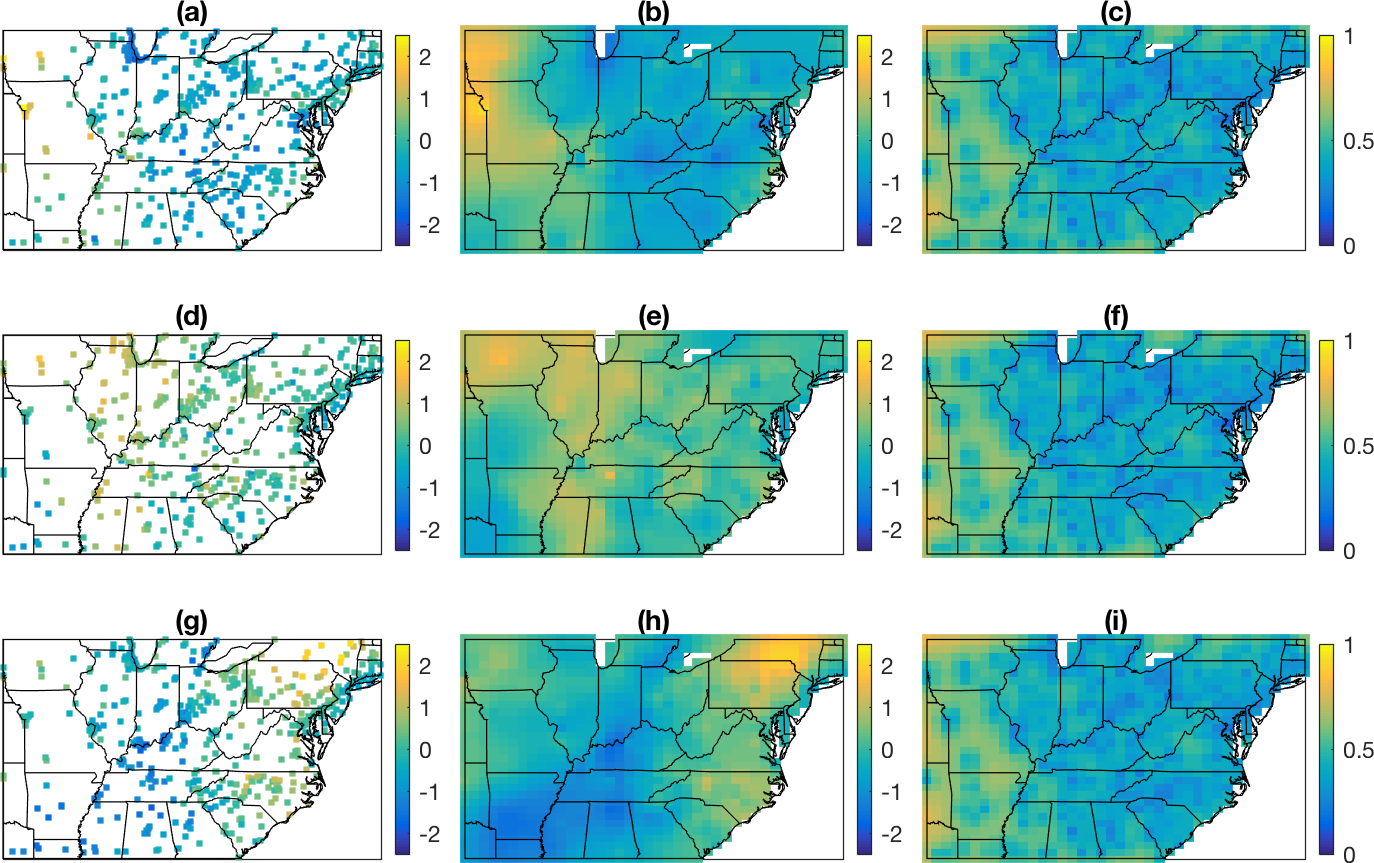}}
\caption{Standardized ozone data and predictions for $Y(\cdot)$ based on 30 by 40 locations on three consecutive days (unit: parts per million). The panels (a), (d), (g) show the standardized ozone on June 14, 15, 16 in 1997. The panels (b), (e), (h) show the posterior mean on June 14, 15, 16 in 1997.  The panels (c), (f), (i) shows the corresponding posterior standard errors.}
\label{fig: predictions for ozone data}
\end{center}
\end{figure*}

Predictions are also carried out over space for different days based on all observed data. Figure~\ref{fig: predictions for ozone data} visualizes the predictions on three consecutive days based on all available data, which clearly shows that AAGP is able to capture the spatio-temporal dependence structures in the data. Predictions from the AAGP model can be used to quantify the impact of ozone pollution once they are processed to the original scale with the seasonal mean added back. Although the time span for this dataset is not latest, this approach can be used for most recent ozone datasets and thus help setup NAAQS for the ozone pollutant.

\section{Discussion} \label{sec: conclusion}
We propose the additive approximate Gaussian process (AAGP) for analyzing large and complex spatio-temporal datasets. It is based on the combination of a CCR nonseparable covariance function and a separable covariance function. The proposed method provides a flexible way to characterize spatio-temporal dependence structures. We also propose a fully conditional Markov chain Monte Carlo algorithm based on the hierarchical representation of the model. This proposed fully Bayesian inference framework allows efficient computation for large spatio-temporal data, and avoids expensive calculation of the marginal likelihood. We show that the new approximation method can give comparable or better prediction results under various scenarios, including different assumptions of covariance functions and a partially observed deterministic function. The analysis of Eastern U.S. ozone data shows that the proposed method can work satisfactorily in terms of predictive accuracy and computational efficiency. 

The proposed  AAGP relies on a CCR covariance function and a separable covariance function, which typically allows fast computation for large spatio-temporal datasets. When the number of data points in space (or time) is large and the number of data points in time (or space) is small, one can also incorporate a purely spatial or temporal CCR method. The proposed fully conditional Markov chain Monte Carlo algorithm can also be applied efficiently. 

 Although we use solely MPP as the CCR component in AAGP, other methods including FSA, NNGP, and MRA can be used alternatively to approximate the nonseparable covariance function, though it requires additional work to investigate how to define and choose the neighbors in space and time efficiently and effectively. This would lead to a more general approximation method. We leave this as an important direction for future research. The corresponding Bayesian inference can be extended to incorporate these approximation methods for the CCR component. The efficiency of the proposed Bayesian inference procedure can be improved further using partially collapsed Gibbs samplers \citep{vanDyk2008, vanDyk2015}. A more rigorous algorithmic development as well as comparison is left for future work. 


\section*{Supplementary Material}
Additional information and supporting material for this article is available online at the journal's website.

\section*{Acknowledgement}
This work was supported in part by an allocation of computing time from the Ohio Supercomputer Center. Ma's research was partially supported by the National Science Foundation under Grant DMS-1638521 to the Statistical and Applied Mathematical Sciences Institute. Any opinions, findings, and conclusions or recommendations expressed in this material are those of the author(s) and do not necessarily reflect the views of the National Science Foundation. Kang's research was supported by the Simons Foundation Collaboration Award (\#317298) and the Taft Research Center at the University of Cincinnati.

\bibliographystyle{apa}
\bibliography{AAGP}
\newpage

\begin{table*}[htbp]
\centering
\normalsize
   \caption{Simulation results from Scenario 1 with a nonseparable space-time correlation function.}
  {\resizebox{1.0\textwidth}{!}{%
  \setlength{\tabcolsep}{.5em}
   \begin{tabular}{l c  c c c c c} 
   \toprule 
    \multicolumn{7}{c}{Scenario 1}  \\  \cline{1-7} \noalign{\vskip 2pt} 
   \multirow{2}{*}{Parameters} & \multirow{2}{*}{True value} & \multirow{2}{*}{Full GP} & \multicolumn{2}{c}{MPP} & \multicolumn{1}{c}{NNGP} & \multicolumn{1}{c}{AAGP}  \\    \noalign{\vskip 1pt}
    & & & $m=250$ &$m=704$ & 15 & $m=250$ \\   \hline \noalign{\vskip 2pt}
$b_1$ & 1 & 0.989(0.980, 0.998) &0.990(0.981, 1.000) &0.993(0.968, 1.019) &1.033(0.717, 1.303)  &0.981(0.959, 1.007) \\
$b_2$ & 0.5 & 0.503(0.473, 0.533) & 0.503(0.491, 0.516) & 0.482(0.445, 0.517) & 0.495(0.382, 0.601) &0.507(0.468, 0.534) \\
$\sigma^2_1$ & 1  &0.978(0.869, 1.096) &1.419(1.236, 1.645) & 1.253(1.070, 1.474)&1.062(0.905, 1.241) &0.869(0.787, 1.038) \\
$\beta$ & 0.8 &  0.841(0.425, 0.994)  &0.935(0.710, 0.999) &0.913(0.549, 0.995) &0.734(0.169, 0.992) & 0.928(0.631, 0.997) \\
$a$ & 1 & 0.991(0.754, 1.255)  &2.070(1.613, 2.659) &2.301(1.728, 3.073) &0.738(0.541, 0.957) & 2.805(2.054, 3.872) \\
$c$ & 5 &  4.768(4.022, 5.710) &5.465(4.503, 6.560) &7.700(6.107, 9.737) &5.302(4.327, 6.573) & 5.542(4.685, 6.750) \\
$\sigma^2_2$ &  &  & & & &0.303(0.221, 0.341) \\
$\phi_s$ & &  &  & & & 0.244(0.035, 0.488) \\
$\phi_t$ & &  &  & & & 1.370(1.344, 1.417)\\
$\tau^2$ & 0.2 & 0.182(0.146, 0.221)  &0.130(0.049, 0.208) & 0.216(0.154, 0.288) &0.189(0.157, 0.219) &0.060(0.021, 0.130) \\
\noalign{\vskip 1.5pt} \hline \noalign{\vskip 1.5pt}
MSPE & & 0.23  &0.51 &0.44 &0.42 & 0.33 \\
ALCI & &1.92 &3.24 &2.90 &2.49 & 2.97 \\ \noalign{\vskip 1.5pt}
\hline \noalign{\vskip 1.5pt}
Time (h) & &23.1 &1.32 &3.92 &3.21 & 2.57 \\ 
 \bottomrule
   \end{tabular}%
   }}
   \label{table: scenario 1} 
\end{table*}

\begin{table*}[htbp]
\centering
\normalsize
   \caption{Simulation results from Scenario 2 with a separable correlation function.}
  {\resizebox{1.0\textwidth}{!}{%
  \setlength{\tabcolsep}{.5em}
   \begin{tabular}{l c  c c c c c c} 
   \toprule 
    \multicolumn{7}{c}{Scenario 2}  \\  \cline{1-7} \noalign{\vskip 2pt} 
   \multirow{2}{*}{Parameters} & \multirow{2}{*}{True value} & \multirow{2}{*}{Full GP} & \multicolumn{2}{c}{MPP} & \multicolumn{1}{c}{NNGP} & \multicolumn{1}{c}{AAGP}  \\   \noalign{\vskip 1pt}
    & & & $m=250$ &$m=704$ & 15 &$m=250$ \\  \hline \noalign{\vskip 2pt}
$b_1$ & 1 &0.989(0.975, 1.004) & 1.003(0.973, 1.033) &0.995(0.962, 1.027) &0.877(0.607, 1.076) &0.989(0.974, 1.003) \\
$b_2$ & 0.5 &0.503(0.483, 0.522) &0.498(0.455, 0.542) &0.493(0.447, 0.539) & 0.490(0.445, 0.598)  &0.503(0.484, 0.522) \\
$\sigma^2_1$ & & &0.083(0.045, 0.163) &0.052(0.027, 0.098)&0.705(0.635, 0.807)  &0.030(0.012, 0.060) \\
$\beta$ & &  &0.737(0.104, 0.987) & 0.718(0.083, 0.989)&0.498(0.026, 0.974) & 0.837(0.372, 0.993)   \\
$a$ & &  &0.768(0.378, 1.564)&1.379(0.679, 2.805)&0.365(0.252, 0.525) & 9.425(2.390, 20.00)  \\
$c$ & & &17.81(9.540, 20.00) &19.08(11.37, 20.00)&7.086(6.267, 8.645)  & 0.198(0.023, 0.835)  \\
$\sigma^2_2$ & 1  &0.953(0.752, 1.195)  & & &  &0.997(0.830, 1.187) \\
$\phi_s$ & 5 &4.968(4.646, 5.213) &  & &  & 5.169(4.922, 5.262) \\
$\phi_u$ & 1 &0.990(0.918, 1.067) &  & &  & 1.005(0.940, 1.083)\\
$\tau^2$ & 0.2 &0.188(0.179, 0.196) &0.950(0.992, 1.037) &1.125(1.072, 1.180) &0.109(0.096, 0.124) & 0.187(0.178, 0.196) \\
\noalign{\vskip 1.5pt} \hline \noalign{\vskip 1.5pt}
MSPE & & 0.02 &0.87 &0.96 &0.23 & 0.02\\
ALCI & &0.58 &0.98 &0.70 &1.88 &0.69 \\ \noalign{\vskip 1.5pt}
\hline \noalign{\vskip 1.5pt}
Time (h) & &40.9 &1.39 &3.94 &3.65 &2.78 \\ 
 \bottomrule
   \end{tabular}%
   }}
   \label{table: scenario 2}  
\end{table*}

\begin{table*}[htbp]
\centering
\normalsize
   \caption{Simulation results from Scenario 3 with a combination of nonseparable and separable correlation functions.}
  {\resizebox{1.0\textwidth}{!}{%
  \setlength{\tabcolsep}{.5em}
   \begin{tabular}{l c  c c c c c c} 
   \toprule 
    \multicolumn{7}{c}{Scenario 3}  \\  \cline{1-7} \noalign{\vskip 2pt} 
   \multirow{2}{*}{Parameters} & \multirow{2}{*}{True value} & \multirow{2}{*}{Full GP} & \multicolumn{2}{c}{MPP} & \multicolumn{1}{c}{NNGP} & \multicolumn{1}{c}{AAGP}  \\    \noalign{\vskip 1pt}
    & & & $m=250$ &$m=704$ &15 & $m=250$ \\   \hline \noalign{\vskip 2pt}
$b_1$ & 1 &0.995(0.944, 0.999)  & 0.969(0.932, 1.007) & 0.965(0.930, 1.001)  &0.822(0.390, 1.209) & 0.969(0.949, 0.997)\\
$b_2$ & 0.5 & 0.502(0.472, 0.532)  & 0.499(0.446, 0.552) & 0.463(0.412, 0.511) &0.490(0.345, 0.612) &0.509(0.475, 0.540) \\
$\sigma^2_1$ & 1 & 0.993(0.863, 1.204)  & 1.615(1.319, 1.957) & 1.915(1.611, 2.240) &2.290(2.245, 2.300) &1.356(1.181, 1.463) \\
$\beta$ & 0.8 &0.847(0.396, 0.994)  & 0.897(0.539, 0.996) &0.850(0.447, 0.994)&0.172(0.005, 0.562) &0.848(0.332, 0.993)\\
$a$ & 1 &0.961(0.705, 1.316)  & 2.379(1.324, 3.677)& 2.172(1.454, 3.223) &3.220(2.714, 4.090) &1.737(1.409, 2.248) \\
$c$ & 5 &4.758(3.857, 6.226)  & 4.973(3.660, 6.639) &3.850(2.885, 5.081) &7.781(6.853, 8.964) &8.253(6.711, 10.43) \\
$\sigma^2_2$ & 1  &0.916(0.730, 1.172) & & &  &0.970(0.806, 1.154)\\
$\phi_s$ & 5 &4.889(4.529, 5.213) & & &  &5.047(4.878, 5.150) \\
$\phi_u$ & 1 &1.003(0.926, 1.071) & & &  &1.037(0.989, 1.095) \\
$\tau^2$ & 0.2 &0.175(0.141, 0.211) & 0.583(0.236, 0.795) & 0.368(0.078, 0.567) &0.504(0.454, 0.559) &0.056(0.019, 0.155) \\
\noalign{\vskip 1.5pt} \hline \noalign{\vskip 1.5pt}
MSPE & &0.28 & 1.40 & 1.30 &0.83 &0.48\\ 
ALCI & &2.10 &3.85 &4.03 &3.52 &3.40 \\ \noalign{\vskip 1.5pt}
\hline \noalign{\vskip 1.5pt}
Time (h) & &106 &1.44 &4.00 &4.34 &3.47 \\ 
 \bottomrule
   \end{tabular}%
   }}
   \label{table: scenario 3} 
\end{table*}

\begin{table*}[htbp]
\centering
\normalsize
   \caption{Cross validation results for standardized ozone from June to August, 1997. Gneiting's nonseparable correlation function is used in all these  three methods, MPP, NNGP, AAGP.  Exponential correlation functions are used for  the separable covariance function in AAGP.}%
  {\resizebox{1.0\textwidth}{!}{%
  \setlength{\tabcolsep}{1.0em}
   \begin{tabular}{l c c c c} 
   \toprule \noalign{\vskip 1.5pt}
   \multirow{2}{*}{Parameters}   & \multicolumn{2}{c}{MPP} & \multicolumn{1}{c}{NNGP} & \multirow{1}{*}{AAGP}  \\   \noalign{\vskip 1.5pt} 
  & $m=490$ & $m=1200$ & 15 & $m=490$ \\ \noalign{\vskip 1.5pt}  \hline \noalign{\vskip 1.5pt} 
$\sigma^2_1$   &0.853(0.796, 0.914) & 0.949(0.874, 1.035) &0.932(0.897, 0.983) & 0.097(0.085, 0.137)  \\
$\beta$ &0.973(0.996, 1.000)  & 0.983(0.907, 0.999) &0.503(0.025, 0.9798) & 0.017(0, 0.101) \\
$a$ (day) &1.674(1.557, 1.782) &1.084(0.992, 1.170)  &1.438(1.285, 1.597) &5.196(3.932, 7.112) \\
$c$ (km) &1508(1406, 1625) & 1311(1202, 1444)  & 388.5(377.1, 412.6) &1999(1996, 2000)  \\
$\sigma^2_2$ & &  &  &0.806(0.756, 0.870)\\
$\phi_s$ (km) &  &  &  &280.1(261.8, 302.9)\\
$\phi_u$ (day) &  &  &  &1.85(1.78, 1.93)\\
$\tau^2$ &0.194(0.187, 0.202) & 0.136(0.132, 0.141) &0.061(0.058, 0.063) & 0.041(0.035, 0.044)\\ \noalign{\vskip 1.5pt}
\hline \noalign{\vskip 1.0pt}
MSPE &0.41 &0.30  &0.16  &0.13\\
ALCI &1.88  & 1.54 &1.44  &1.05 \\
\hline \noalign{\vskip 1.5pt}
Time (h) &41.8 &91.8 &150 &87.7 \\
 \bottomrule
   \end{tabular}%
   }}
   \label{table: cross validation for Jun-Aug} 
\end{table*}

\end{document}